# 5G Waveforms for Overlay D2D Communications: Effects of Time-Frequency Misalignment

Quentin Bodinier*, Arman Farhang†, Faouzi Bader*, Hamed Ahmadi‡, Jacques Palicot*, and Luiz A. DaSilva†
*SCEE/IETR - CentraleSupelec, Rennes, France,
†CONNECT - Trinity College Dublin, Ireland,
‡University College Dublin, Ireland.
Email : quentin.bodinier@supelec.fr

***Abstract*—This paper analyses a scenario where a Device-To-Device (D2D) pair coexists with an Orthogonal Frequency Division Multiplexing (OFDM) based incumbent network. D2D transmitter communicates in parts of spectrum left free by cellular users, while respecting a given spectral mask. The D2D pair is misaligned in time and frequency with the cellular users. Furthermore, the D2D pair utilizes alternative waveforms to OFDM proposed for 5G. In this study, we show that it is not worth synchronising the D2D pair in time with respect to the cellular users. Indeed, the interference injected into the incumbent network has small variations with respect to time misalignment. We provide interference tables that encompass both time and frequency misalignment. We use them to analyse the maximum rate achievable by the D2D pair when it uses different waveforms. Then, we present numerical results showing what waveform should be utilized by the D2D pair according to the time-frequency resources that are not used by the incumbent network. Our results show that the delay induced by linearly convolved waveforms make them hardly applicable to short time windows, but that they dominate OFDM for long transmissions, mainly in the case where cellular users are very sensitive to interference.**

## I. Introduction

The advent of Device-To-Device (D2D) communication as a new application in $5^{\text{th}}$ Generation wireless networks (5G) raises a number of challenges. In particular, coexistence of D2D pairs with cellular users may affect the performance observed by the latter. To avoid that, a first standard for D2D communication has been proposed [1], requiring consequent interaction between the base station and the D2D pairs.

With increasing number of D2D links, perfect synchronization between those and the native cellular subscribers is infeasible. Therefore, it is important to consider schemes where D2D users are not perfectly synchronized both in time and frequency with respect to the cellular users. In that case, it is well known that Orthogonal Frequency Division Multiplexing (OFDM) suffers from very high multiple user interference when timing and frequency offsets go beyond certain values [2], [3]. As a consequence, the overall quality of transmission seen by cellular users may decrease dramatically.

To cope with that challenge, a number of solutions based on both power loading and waveform design have been proposed [4]–[14]. On one hand, to the best of our knowledge, studies of interference caused by time-frequency misalignment are restricted to the case where the misaligned user has the same waveform as the others [3], [15], and there is no work addressing the interference caused by a given waveform onto OFDM. On the other hand, power-loading studies rely on interference models which do not take time-frequency misalignment into account [4]–[7]. However, accurately estimating the caused interference is of high importance, as D2D users aim to maximise their data rate subject to constraints of total power and maximum injected interference.

Therefore, it is crucial to analyse the effects of time-frequency misaligment on the interference caused by D2D pairs onto cellular users. In this paper, we derive interference tables that can be used to find the optimal solution of the power loading problem for D2D users following the model of [6]. These results allow us to analyse the performance of a D2D pair as a function of the waveform they utilise.

As a matter of fact, the rate achievable by D2D users is highly dependent on their air interface. Indeed, utilization of appropriate waveforms can limit Out Of Band (OOB) emissions of D2D users and therefore allows them to allocate higher power to their active subcarriers. It is well known that OFDM suffers from large sidelobes in frequency due to its rectangular window in time [11], [16], [17]. To overcome this problem, numerous new waveforms with intrinsic filtering properties have emerged [7]–[14]. These waveforms all rely on Filter Bank Multi Carrier (FBMC) modulation. In this work, we focus on four recently proposed FBMC based waveforms, namely Filtered Multi-Tone (FMT), Offset Quadrature Amplitude Modulated-Orthogonal Frequency Division Multiplexing (OFDM/OQAM), Lapped FBMC and GFDM.

In FMT modulation [8], [9], the assumption is that there is no overlapping between adjacent subcariers. Therefore, FMT suffers from some bandwidth efficiency loss. To increase spectral efficiency, OFDM/OQAM [10], [11] allows for adjacent subcarriers to overlap. Differently from FMT, real symbols having Pulse Amplitude Modulation (PAM) are transmitted on each subcarrier. A $\frac{\pi}{2}$ phase difference is applied to adjacent subcarriers, which provides real-domain orthogonality. However, OFDM/OQAM requires doubling the symbol rate. To avoid that, Lapped FBMC modulation has been recently proposed by Bellanger et al. in [12]. In this scheme, the number of subcarriers is doubled instead of the number of time symbols.

A drawback common to all the aforementioned modulations is the transient imposed by their transmit and receive filters. Generalized Frequency Division Multiplexing (GFDM) tackles this problem by application of cyclic pulse shaping [13]. However, this comes at the expense of higher OOB emissions, which is due do the discontinuities induced in the signal by truncation in time with a rectangular window [14].

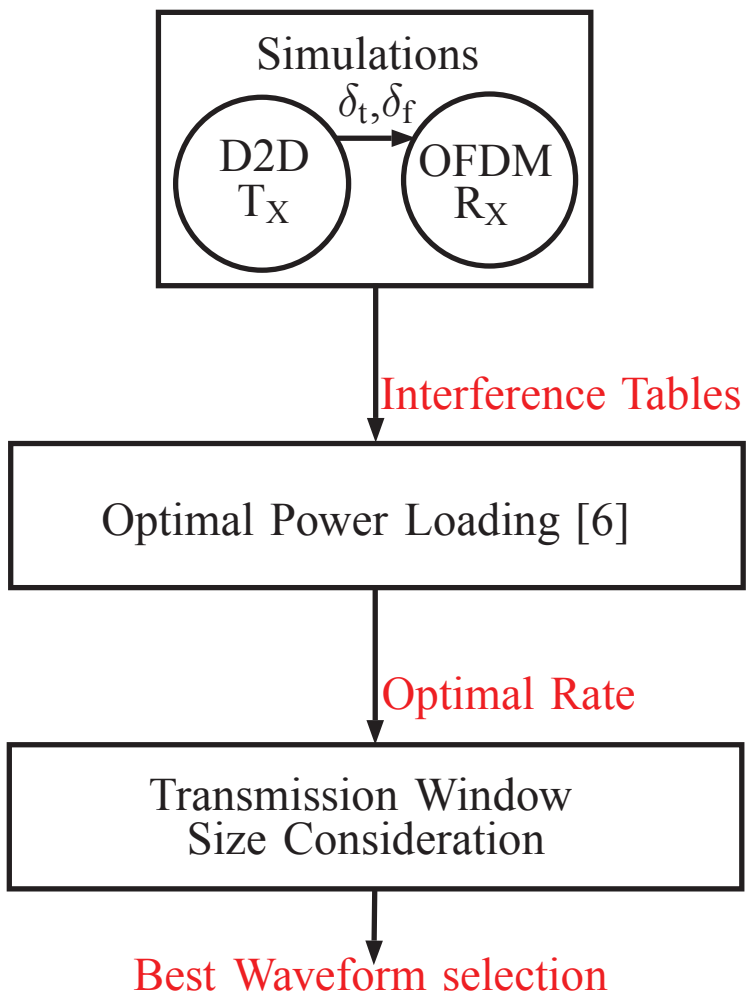

Fig. 1: Paper Synopsis

In this paper, we analyse time-frequency misalignment effects of the D2D pair on the incumbent network. We evaluate the interference injected by the D2D pair on the incumbent network for each waveform separately. This allows us to build interference tables to be used as inputs to the power loading solution and analyse the maximum rate achievable through each waveform. We consider different resource sizes in time and frequency and calculate the resulting data rate. This analysis reveals the strength and weaknesses of different waveforms to be used by the D2D pair as a function of available time-frequency resources. Our approach is summarized in Fig. 1.

The remainder of this paper is organized as follows. In section II, the system model and the analysed waveforms are presented. Section III is dedicated to the interference analysis, whereas section IV presents the rate optimization problem and its solution. Section V presents simulation results. Finally, section VI concludes this paper.

## II. Considered Scenario and Studied Waveforms

### A. Considered scenario

In this study, we consider an OFDM based incumbent network with subcarrier spacing $\Delta F$ where a certain group of subcarriers is free and can be utilized by a D2D pair. It is assumed that the base station broadcasts information about its unused time-frequency resources. This information includes the number of free subcarriers $M_{\text{f}}$ and the number of OFDM symbols during which the band will be free $N_{\text{f}}$. We define $B_{\text{i}}$ as the band ocuppied by the cellular users, $B_{\text{f}}$ the free band used by the D2D pair, and $I_{\text{th}}$ as the interference threshold for $B_{\text{i}}$. It is assumed that the D2D pair can misalign its transmission frequency with respect to $B_{\text{f}}$. Besides, we assume that, even though the D2D transmission is contained in the duration of the opportunity, time synchronism between D2D and cellular users at the symbol level may not be achieved. We denote $\delta_{\text{t}}$ and $\delta_{\text{f}}$ as the offsets in time and frequency. The scenario is presented in figure 2, where the grey grid represents the time-frequency resources of the incumbent network. These resources follow the 3GPP Long Term Evolution (LTE) standard.

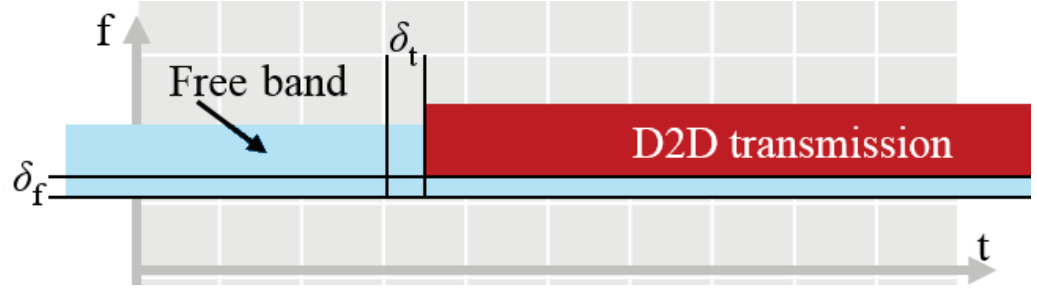

Fig. 2: Time-Frequency layout of the situation with time-frequency misplacement representation.

### B. Waveforms under consideration

To pave the way for the analysis conducted in Section III, we briefly explain the mathematical representation of the waveforms under study to be used by the D2D pair.

*1) FMT:* In this scheme, complex Quadrature Amplitude Modulation (QAM) symbols are linearly pulse shaped using the prototype filter $\mathbf{g} = [g[0] \dots g[KP-1]]^{\text{T}}$, where $K$ is the overlapping factor and $P$ is the number of samples per time symbol. $M$ is the number of subcarriers, $N$ the number of symbols in time, and $\mathbf{d}_m = [d_m[0] \dots d_m[N-1]]^{\text{T}}$ the vector of modulated symbols on the $m$th subcarrier. Finally, the FMT signal can be obtained as [9]

$$x_{\text{FMT}}[k] = \sum_{n=0}^{N-1} \sum_{m=0}^{M-1} d_m[n] g[k-nP] e^{j2\pi \frac{m}{M} k}, \quad \forall k = 0 \dots (N+K-1)P-1. \tag{1}$$

*2) OFDM/OQAM:* In this modulation format, the signal can be derived in a similar way to FMT, where $P = M$, $\{d_m\}$ symbols are drawn from a PAM constellation, and a phase factor $\theta_m[n] = e^{j\frac{\pi}{2}\lfloor \frac{n+m}{2} \rfloor}$ is introduced where $n$ and $m$ the time-frequency indexes respectively. Besides, subsequent symbols are separated by $\frac{M}{2}$ samples in time, which implies doubling the symbol rate. Therefore, the transmit signal can be expressed as [18]

$$x_{\text{OQAM}}[k] = \sum_{n=0}^{N-1} \sum_{m=0}^{M-1} \Big( d_m[n] \theta_m[n] g[k - n\frac{M}{2}] \times e^{j2\pi \frac{m}{M}(k - \frac{KM-1}{2})} \Big), \quad \forall k = 0 \dots (P + K - \frac{1}{2})M - 1. \tag{2}$$

*3) Lapped FBMC:* This scheme is in essence similar to OFDM/OQAM systems. However, for this modulation the symbol rate is not doubled. As defined in [12], the emitted signal can be written as

$$x_{\text{Lapped}}[k] = \sum_{n=0}^{N-1} \sum_{m=0}^{M-1} d_m[n] g[k-nM] e^{j(k-\frac{1}{2}+\frac{M}{2})(m-\frac{1}{2})\frac{\pi}{M}}, \quad \forall k = 0 \dots (P+1)M-1. \tag{3}$$

It is worth mentioning that the used filter is not tunable and is written as

$$g_{\text{Lapped}}[k] = -\sin\left[(k-\frac{1}{2})\frac{\pi}{2M}\right], \forall k = 0\dots 2M-1. \quad (4)$$

*4) GFDM:* The data is packed in blocks of $N_{\text{b}}$ symbols and a Cyclic Prefix (CP) is added to the beginning of each block. Every symbol is circularly convolved with a time shifted version of the same circular filter $\mathbf{g}$. Denoting as mod the modulo operation, the signal corresponding to one GFDM block is expressed as follows [14]

$$x_{\text{GFDM}}[k] = \sum_{n=0}^{N_{\text{b}}-1}\sum_{m=0}^{M-1} d_m[n] g\left[(k-nM)\text{mod}(N_{\text{b}}M)\right] e^{j2\pi k\frac{m}{M}}, \quad \forall k = 0\dots MN_{\text{b}}-1. \quad (5)$$

## III. Interference Analysis

To assign the optimal power distribution to the D2D subcarriers, it is vital to know how much power leaks to the adjacent band depending on which waveform is utilized by the D2D pair. The classical way to compute the leakage is to integrate the Power Spectral Density (PSD) of the interfering signal on the band that suffers from the interference. However, this model does not take into account the time window of the incumbent. This is of paramount importance as the incumbent only considers a time window with a specific width based on its own parameters. Besides, it has been shown in [19] that PSD is not a suitable measure to evaluate the inter-system interference for the scenario of interest to this paper. We therefore employ the instantaneous interference model proposed in [19] to compute the interference $I_m^l$ injected to the $l$th subcarrier of the incumbent by the D2D signal $x_m$ where only subcarrier $m$ is utilized. This allows us to scrutinize the interference injected by each individual subcarrier to the incumbent network. Note that the same subcarrier spacing is used for both the D2D pair and the incumbent network. Therefore, we have

$$I_m^l = \int_{t=0}^{T} \left|\left(g_r^l * x_m\right)(t)\right|^2 dt, \quad (6)$$

where $g_r^l$ is the receiver filter on subcarrier $l$ and $*$ denotes the convolution operation. As the receiver suffering from interference is based on OFDM, in the discrete time domain, (6) becomes

$$I_m^l = \sum_{n=0}^{N-1} \frac{T_{\text{s}}}{M+N_{\text{CP}}} \sum_{k=n(M+N_{\text{CP}})+N_{\text{CP}}}^{(n+1)(M+N_{\text{CP}})} \left|x_m[k]e^{-j2\pi k\frac{l}{M}}\right|^2, \quad (7)$$

where $N$ is the total number of OFDM symbols corresponding to the time span $T$, $N_{\text{CP}}$ is the number of CP samples and $M$ the number of samples per OFDM symbol of length $T_{\text{s}}$. According to the signal models presented in the former section, for any of the analysed waveforms, $x_m$ can be written as

$$x_m[k] = a_m[k]e^{j2\pi k\frac{m}{M}}, \forall k = 0\dots N(M+N_{\text{CP}})-1, \quad (8)$$

where $a_m$ encompasses the modulated signal on subcarrier $m$. Therefore, putting (8) in (7) and taking into account timing and frequency offsets between the interferer and the receiver, we have

$$I_m^l(\delta_{\text{t}},\delta_{\text{f}}) = \sum_{n=0}^{N-1}\left(\frac{T_{\text{s}}}{M+N_{\text{CP}}}\sum_{k=n(M+N_{\text{CP}})+N_{\text{CP}}}^{(n+1)(M+N_{\text{CP}})} \left|a_m[k]e^{j2\pi\left(k\frac{m-l+\delta_{\text{f}}}{M}+\frac{\delta_{\text{t}}}{M}\right)}\right|^2\right). \quad (9)$$

Then, the mean interference seen at subcarrier $l$ at the receiver of the incumbent can be written as

$$I_{m_{\text{mean}}}^l(\delta_{\text{t}},\delta_{\text{f}}) = \frac{1}{N}\mathbb{E}_{d_m}(I_m^l(\delta_{\text{t}},\delta_{\text{f}})), \quad (10)$$

where $\mathbb{E}_{d_m}$ represents the expectation with respect to the symbols transmitted on subcarrier $m$. We point out that when a larger number of subcarriers are utilized by the D2D pair, the total interference is equal to the sum of the interference caused by each subcarrier.

Finally, the total interference injected by the D2D transmission on the free band $B_{\text{f}}$ to the incumbent band $B_{\text{i}}$ is

$$I_{B_{\text{i}}}^{B_{\text{f}}} = \sum_{m\in B_{\text{i}},\ l\in B_{\text{f}}} \frac{P_m}{P_0} I_{m_{\text{mean}}}^l(\delta_{\text{t}},\delta_{\text{f}}), \quad (11)$$

where $P_m$ is the power assigned to the $m$th subcarrier and $P_0$ is a reference value of $1W$. This method can be used to compute the interference tables for the different analysed waveforms.

## IV. D2D Pair Power Optimization

The interference tables that are derived in the previous section can be used as input to a power optimization block to maximize the rate of D2D users.

### *A. Power Optimization Problem*

In the scenario of interest to this paper, the D2D transmitter has to optimize its power allocation to maximize its rate while satisfying the total power budget constraint, $P_{\text{t}}$, and maximum injected interference $I_{\text{th}}$ to the incumbent band $B_{\text{i}}$. As shown in the previous section, interference injected onto cellular users depends on both the frequency and time misalignments of the D2D pair with respect to the incumbent network. However, it is assumed that the D2D pair is only aware of $\delta_{\text{t}}$ and $\delta_{\text{f}}$ ranges. $\delta_{\text{t}} \in \left[-0.5(\text{T}_{\text{s}}+\text{T}_{\text{CP}}), 0.5(\text{T}_{\text{s}}+\text{T}_{\text{CP}})\right)$ and $\delta_{\text{f}} \in \left[-\delta_{f_{\max}}, \delta_{f_{\max}}\right]$ where $\delta_{f_{\max}}$ is the maximum misalignment in frequency.

In [19], the mean value of interference caused by $\delta_{\text{t}}$ is considered. However, a more stringent policy is required to fully protect the cellular users, as the interference caused by the D2D pair may take higher values than its mean. Therefore, we propose to perform the power allocation considering the maximum value for the injected interference. Hence, we define the maximum interference factor from a given subcarrier, $m$, to the incumbent band as

$$\Omega_m = \sum_{l\in B_{\text{i}}} \frac{\max_{\delta_{\text{f}},\delta_{\text{t}}} I_{m_{\text{mean}}}^l(\delta_{\text{t}},\delta_{\text{f}})}{P_0}. \quad (12)$$

TABLE I: Number of data symbols useful to D2D pairs

| **Waveform** | Useful D2D Symbols |
|---|---|
| **OFDM** | $N_{\text{f}}$ |
| **FMT** | $N_{\text{f}} - K + 1$ |
| **OFDM/OQAM** | $\lfloor N_{\text{f}} \frac{M+N_{\text{CP}}}{M} - K + \frac{1}{2} \rfloor$ |
| **Lapped FBMC** | $\lfloor N_{\text{f}} \frac{M+N_{\text{CP}}}{M} - 1 \rfloor$ |
| **GFDM** | $N_b \lfloor \frac{N_{\text{f}}}{N_{\text{b}}} \frac{N_{\text{b}}(M+N_{\text{CP}})}{N_{\text{b}}M+N_{\text{CP}}} \rfloor$ |

On this basis, we define the following optimization problem for D2D pairs.

$$
\begin{aligned}
P1 : \max_{P_m} & \sum_{m=0}^{M_{\text{f}}-1} \log_2(1 + \frac{P_m^2}{\sigma_{\text{N+I}}^2}), \\
\text{s.t.} & \\
& \sum_{m \in B_{\text{f}}} P_m \Omega_m, \leq I_{\text{th}}, \\
& \sum_{m=0}^{M_{\text{f}}-1} P_m \leq P_{\text{t}}, \\
& P_m \geq 0, \forall m \in \{1 \dots M_{\text{f}}\}.
\end{aligned} \tag{13}
$$

where $\sigma_{\text{N+I}}^2$ is the term accounting for both white noise and interference coming from cellular users.

After deriving the Karush-Kuhn-Tucker (KKT) conditions for the optimization in (13), the optical power allocation on subcarrier m is given by [6] as

$$
P_m^* = \max \left( 0, \frac{1}{\alpha \Omega_m + \beta} - \sigma_{\text{N+I}}^2 \right), \tag{14}
$$

$\alpha$ and $\beta$ being the Lagrangian coefficients relative to (13).

### *B. Considering Transmission Window Duration*

As a static interference constraint is assumed during the whole D2D transmission, we can simply compute the total number of bits transmitted as

$$
b = T_{\text{useful}} * \sum_{m=0}^{M_{\text{f}}-1} \log_2(1 + \frac{P_m^*}{\sigma_{\text{N+I}}^2}), \tag{15}
$$

where $T_{\text{useful}}$ is the duration in which useful symbols can transmitted. This value depends on the utilized waveform. Indeed, for filter banks with linear pulse shaping like FMT and Lapped FBMC, the overlapping factor $K$ introduces a delay of $K - 1$ symbols in the time domain. For OFDM/OQAM, the delay imposed by the transmit and receive filters is $K - \frac{1}{2}$ symbols in time as symbols are separated by $\frac{T_s}{2}$. In contrast, OFDM and GFDM do not suffer from any delay. In fact, the block structure of GFDM brings some limitations, as the length of the whole block is fixed for any number of active symbols. Thus, the transmission window can only be fully utilized in time if its duration is a multiple of the GFDM block length. Table I presents the number of usable symbols for each waveform as a function of both D2D and incumbent parameters during a transmission window of length $N_{\text{f}}$ symbols.

TABLE II: Proposed D2D Waveform Parameters

| **Waveform** | Samples per symbol | CP samples | Filter | Overlapping factor. |
|---|---|---|---|---|
| **OFDM** | $M$ | $N_{\text{CP}}$ | Irrelevant | Irrelevant |
| **FMT** | $M + N_{\text{CP}}$ | 0 | RRC, rolloff 0.2 | 6 |
| **OFDM/ OQAM** | $M$ | 0 | Phydias [18] | 4 |
| **Lapped FBMC** | $M$ | 0 | Sine [12] | 2 |
| **GFDM** | $M$ | $N_{CP}$ | RRC, rolloff 0.2 | 5 |

## V. Numerical Results

In this section, we present a broad set of numerical results analyzing the effects of time-frequency misalignment of the D2D pair and performance of the optimal power allocation scheme discussed in Section IV for different waveforms.

### *A. System Setup*

We consider an incumbent system following similar parameters to 3GPP LTE standard. The OFDM cellular user occupies 180 subcarriers, which corresponds to 15 LTE resource blocks along the frequency axis. Besides, it uses $M = 180$ samples per symbol and $N_{\text{CP}} = 12$ samples. The length of the transmission window in time varies from $N_{\text{f}} = 1$ to 100 OFDM symbols. In the center of the incumbent band, a free band $B_{\text{f}}$ divided into $M_{\text{f}}$ subcarriers of 15 kHz is unused. This band is utilized by the D2D pair. No guard band is considered in this study. The parameters of each waveform under study for utilization by the D2D pair are listed in Table II. RRC refers to the Root Raised Cosine filter.

### *B. Computation of Interference Tables*

In this subsection, we compute the interference caused by the D2D transmission according to (10). Fig. 3 illustrates the interference injected by one active subcarrier on the incumbent band as a function of $\delta_{\text{t}}$ when $\delta_{\text{f}} = 0$. We notice that OFDM based D2D transmission does not interfere at all if the timing offset is contained within the CP. However, when $\delta_{\text{t}}$ falls outside the CP, there is a big increase in the amount of interference to the incumbent band. On the other hand, the interference caused by other waveforms does not have a high variation with respect to $\delta_{\text{t}}$. This result reveals that it may not be worth synchronising D2D transmission in time with the cellular users when different waveforms are utilised by the D2D pair.

Fig. 4 shows the interference that is injected by an active subcarrier with unitary power on 20 neighboring OFDM subcarriers as a function of $\delta_{\text{t}}$ for different waveforms. Interference caused by Lapped FBMC and OFDM/OQAM is weakly affected by the timing offset. GFDM and FMT show slight variations with respect to $\delta_{\text{t}}$. On the contrary, the figure for OFDM shows high variations along the $\delta_{\text{t}}$ axis.

In addition, we evaluate the interference that is caused by the different waveforms. We present the mean and maximum interference with respect to $\delta_{\text{t}}$ in Figs. 5a and 5b.

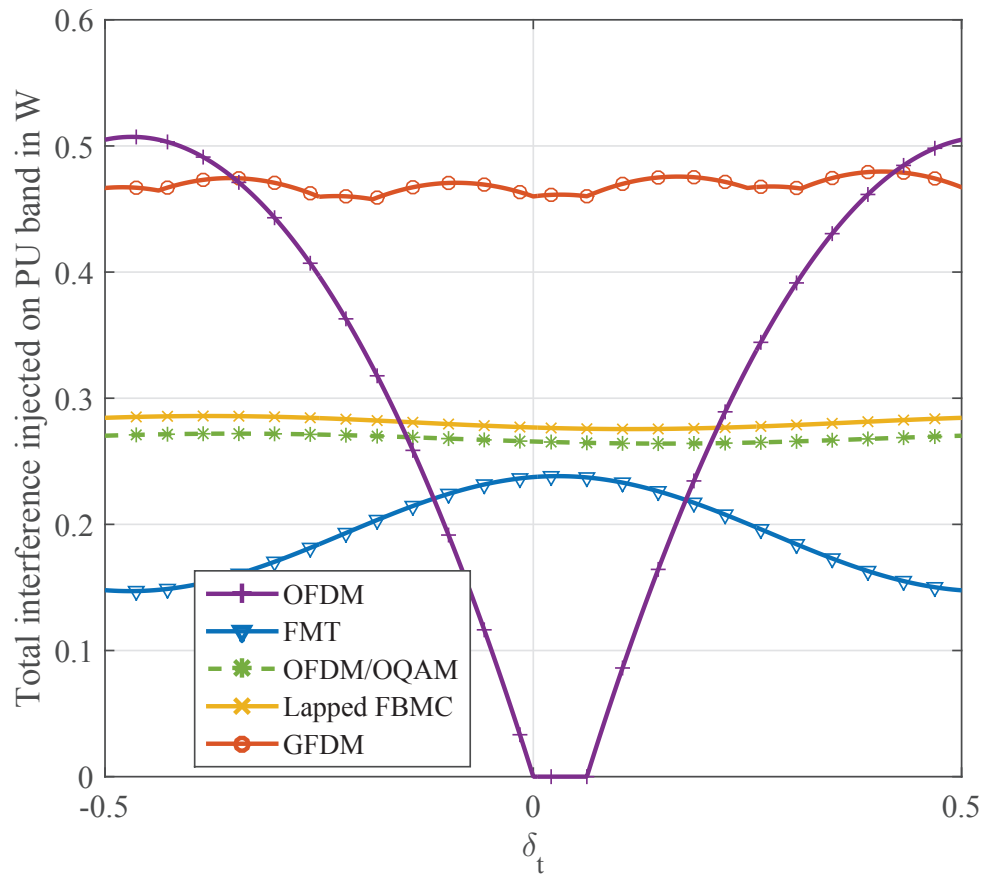

Fig. 3: Total injected interference on the incumbent band as a function of $\delta_t$. Timing offset has limited impact on injected interference except in the case where D2D transmitters use OFDM.

Our observations are threefold: First, it appears that OFDM is the only waveform that shows a significant difference between its mean and maximum injected interference on the OFDM based incumbent. Other waveforms show a difference of approximately 0 to 1dB. Second, if mean interference is considered, GFDM causes the highest interference. However, if the maximum interference is considered, OFDM based D2D pair has the worst performance. Third, we point out that values of interference injected by 5G waveforms are surprisingly high. For OFDM/OQAM for example, the PSD based model predicts an attenuation of $-60$ dB at subcarrier distance of 2 [19], whereas our interference tables show that the interference power seen by an OFDM receiver at subcarrier distance of 2 is $-18.5$ dB. This is due to the fact that the OFDM demodulator performs Fast Fourier Transform (FFT) on a time window that may be much shorter than the length of the symbol of the other waveform. Therefore the signal suffers from discontinuities that produce projections on the whole incumbent spectrum.

Finally, we point out that interference tables are presented for $\delta_f = 0$. However, as $\delta_f$ only acts as a frequency shift in (10), when $\delta_f \neq 0$, the interference can be directly taken from interference tables by taking the corresponding subcarrier distance into account. As a case in point, if the subcarrier distance is $-2$ and $\delta_f = 1$, interference value corresponding to an actual subcarrier distance of $-1$ should be read from the table.

### C. Transmission Performance

In this subsection, we consider the total amount of data that can be transmitted by the D2D pair during the transmission window as a measure to evaluate the performance of different waveforms. We use the interference tables derived in the previous subsection to calculate the total amount of data that can be transmitted during the transmission window based on the waveform that is utilised. We consider a transmission band consisting $M_f = 12$ free subcarriers, an interference constraint $I_{th}$ of either 1 W or 1 mW, and variable number of

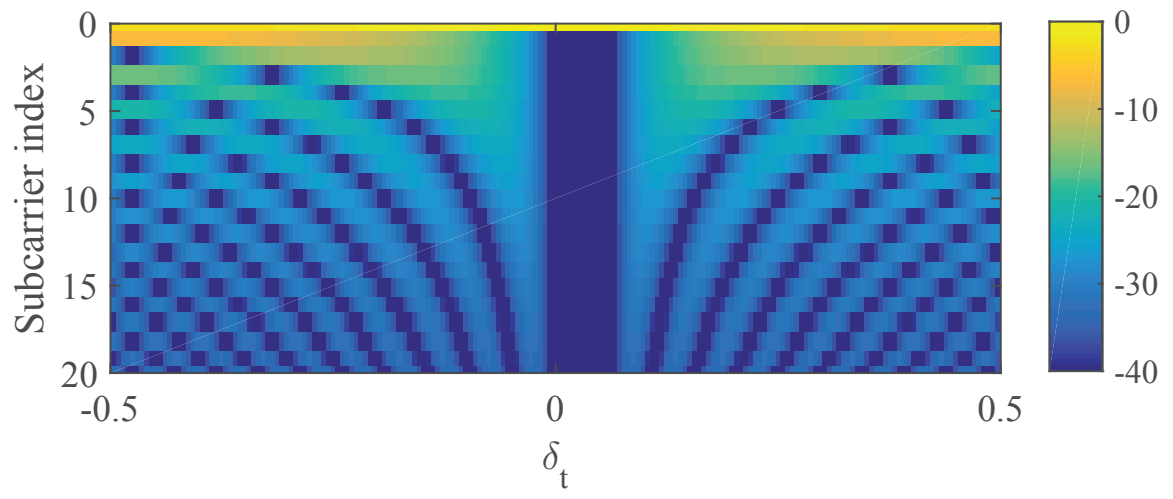

(a) OFDM

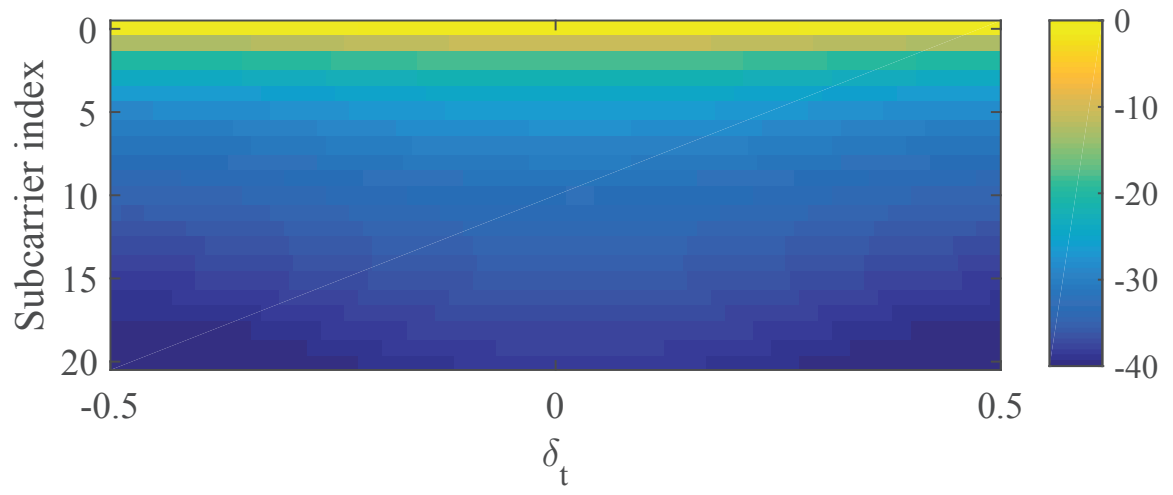

(b) FMT

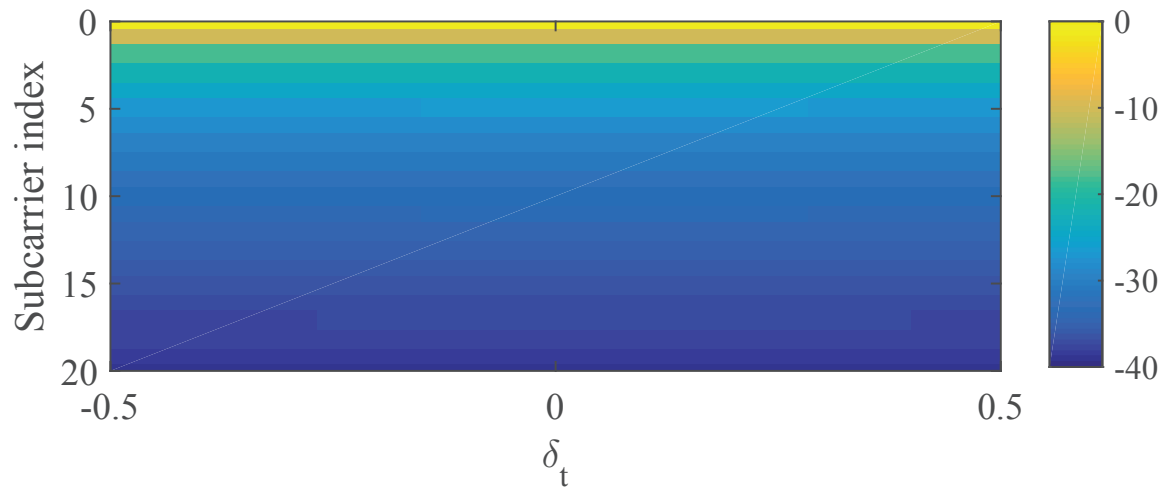

(c) OFDM/OQAM

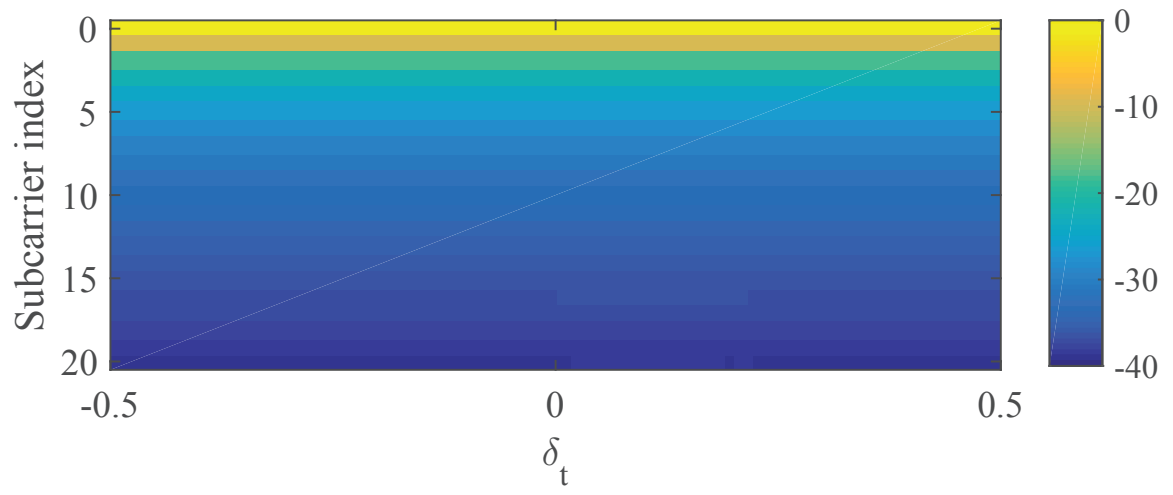

(d) Lapped FBMC

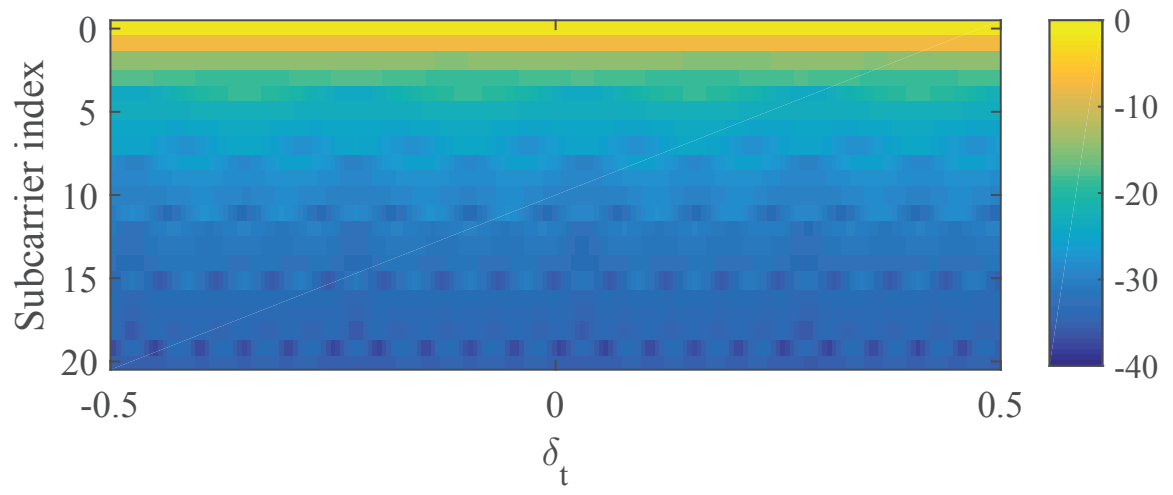

(e) GFDM

Fig. 4: Interference in dB caused by an active D2D at subcarrier 0 on 20 neighboring OFDM subcarriers of the incumbent in function of $\delta_t$ for different waveforms. Only OFDM shows a figure significantly varying along the $\delta_t$ axis. For the other studied waveforms, perfect time synchroization does not significantly reduce injected interference.

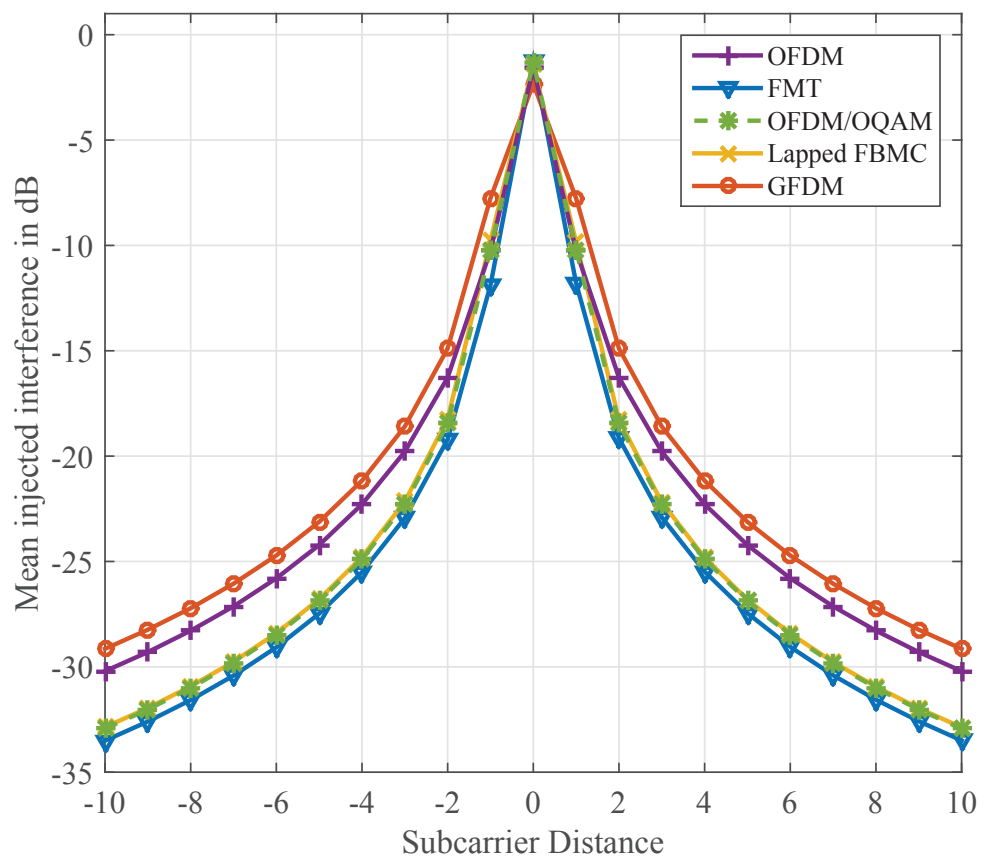

(a) Mean Interference Table on OFDM receiver

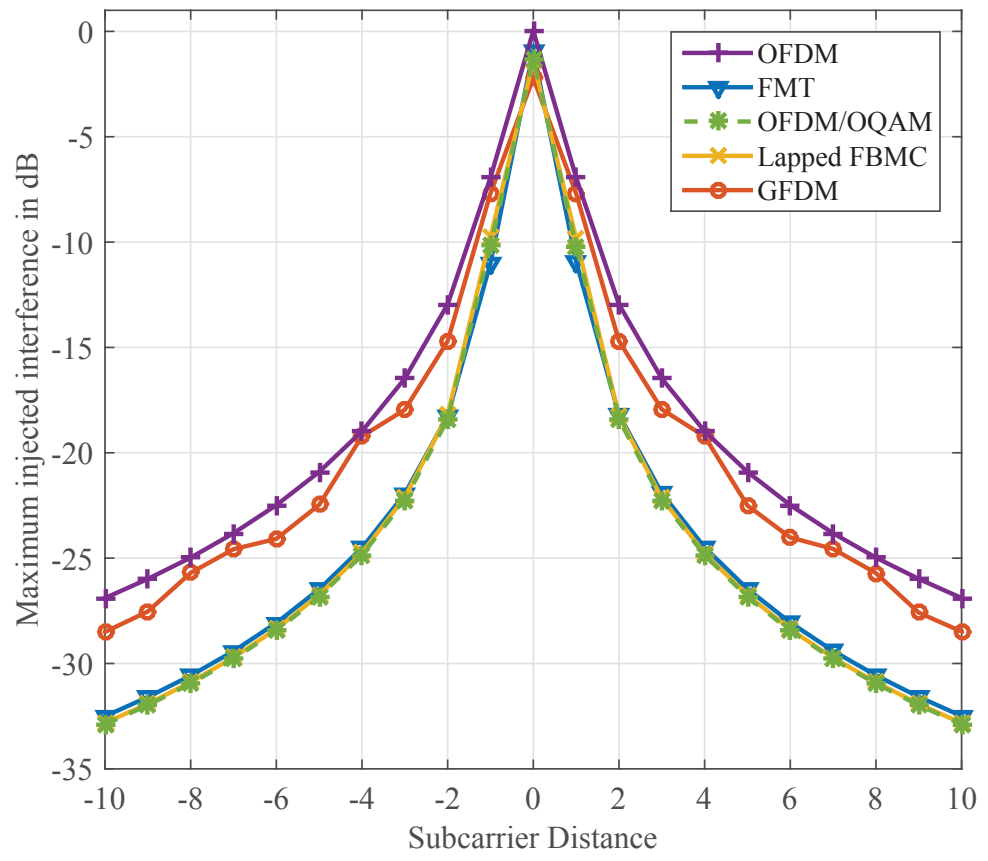

(b) Maximum Interference Table on OFDM receiver

Fig. 5: Maximum and Mean interference seen by a subcarrier of OFDM receiver as a function of its distance to an active 5G waveform subcarrier.

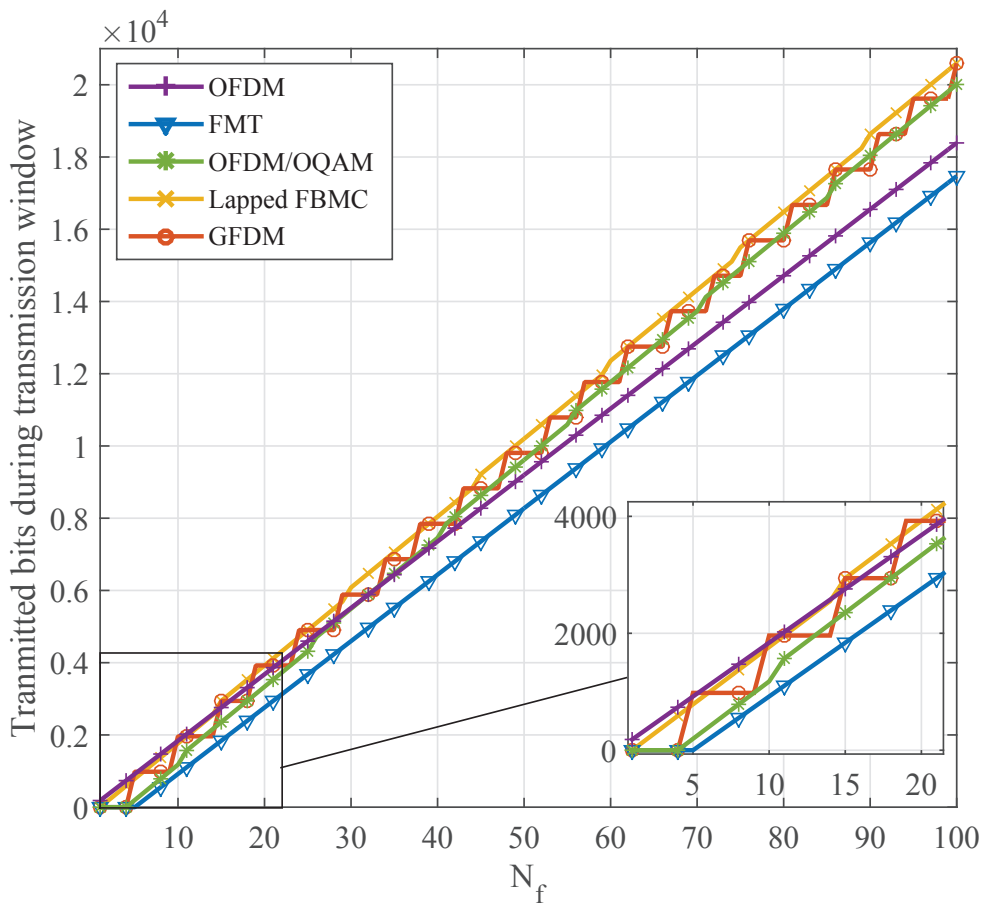

(a) $I_{\text{th}} = 1W$

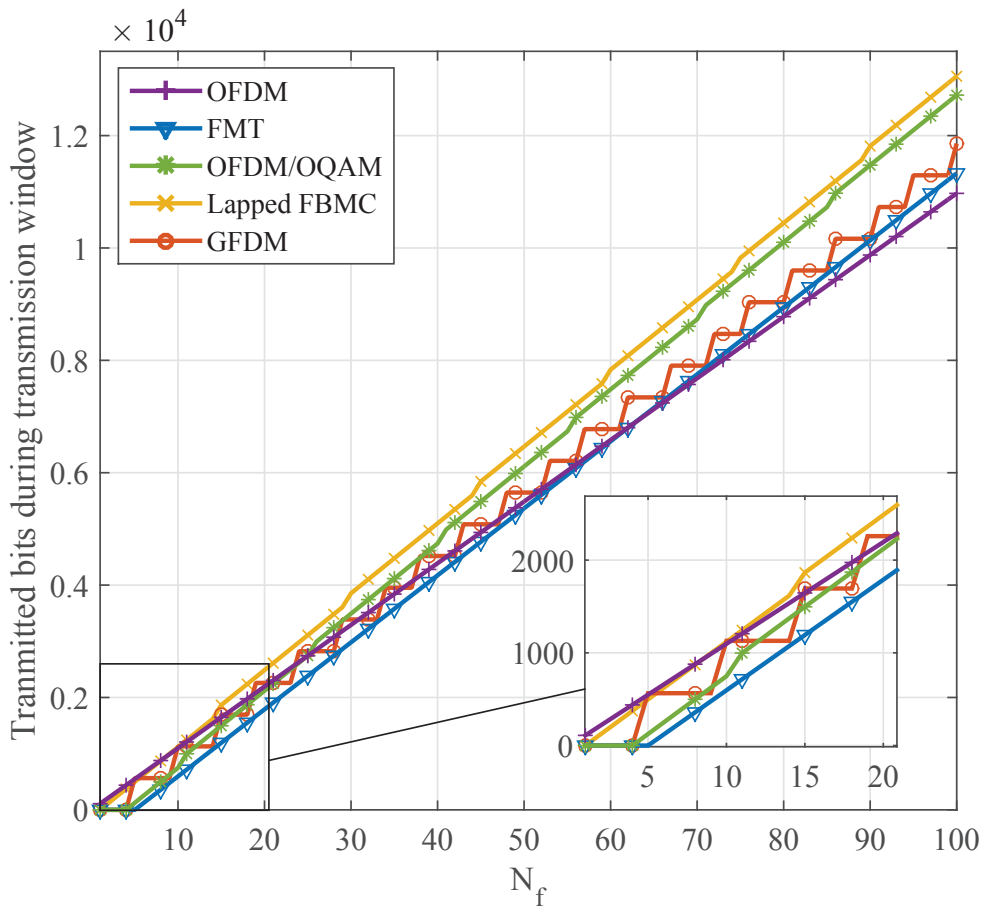

(b) $I_{\text{th}} = 1mW$

Fig. 6: Bits transmitted as a function of available OFDM time symbols in the transmission window

time symbols $N_{\text{f}} \in [1, 100]$. Besides, the maximum frequency misalignment is one subcarrier spacing (i.e. $\delta_{\text{f}_{\text{max}}} = 1$). Note that, $\sigma^2_{\text{N+I}}$ is assigned a low value of $10^{-6}$ to keep this paper focused on the interference injected by D2D pairs onto cellular users. The amount of data transmitted by the D2D pair as a function of $N_{\text{f}}$ is obtained for two values of $I_{\text{th}}$ in Fig. 6.

The presented results bring insight into which waveform performs best for different time window lengths and interference constraint. It seems that for transmission windows shorter than 10 symbols, OFDM is the best choice, as it does not suffer from any transmission delay. Therefore, linearly pulse shaped waveforms can compete only when the transmission window starts getting wider than 10 OFDM symbols. It can be noticed that Lapped FBMC shows a promising performance. This is due to the fact that it has a very short delay about only one symbol, and injects interference comparable to that of OFDM/OQAM and FMT. FMT suffers from a large delay during transmission and seems not to be an appropriate candidate for low latency applications. However, OFDM/OQAM performance stays very close to the Lapped FBMC. Interestingly, the performance of OFDM starts to degrade for time windows of width larger than $N_{\text{f}} = 15$ regardless of the interference constraint. This is the result of its spectral efficiency loss due to the presence of a CP. Note that, even though GFDM seems to be a potential competitor to OFDM/OQAM and Lapped FBMC when the interference constraint is very relaxed, it cannot efficiently cope with stringent interference constraints. This is the consequence of its high interference leakage as shown in Fig. 5.

Finally, we present results corresponding to a scenario where the time-frequency window is equal to 1 LTE Time Transmission Interval (TTI) and 12 subcarriers in frequency. This transmission window is indeed very short and hence waveforms with linear pulse shaping may suffer from the delay imposed by the transient of their transmit and receive filters. Fig. 7 depicts the performance of different waveforms as a function of the interference constraint. All the waveforms show a similar behaviour in which the number of transmitted bits saturate after a certain value of $I_{\text{th}}$. This corresponds to the point where the total power budget becomes the dominating factor to consider in (13). Furthermore, it can be seen once again that Lapped FBMC achieves the best performance as it offers a good trade-off between latency and interference. However, for $I_{\text{th}} \geq -10$ dB, OFDM achieves the best performance as the interference constraint is not restrictive anymore. It is worth mentioning that a specific number of symbols can be transmitted during a TTI corresponding to each waveform (see table I). In particular, GFDM performance is limited because it can only fit 10 symbols in the transmission window.

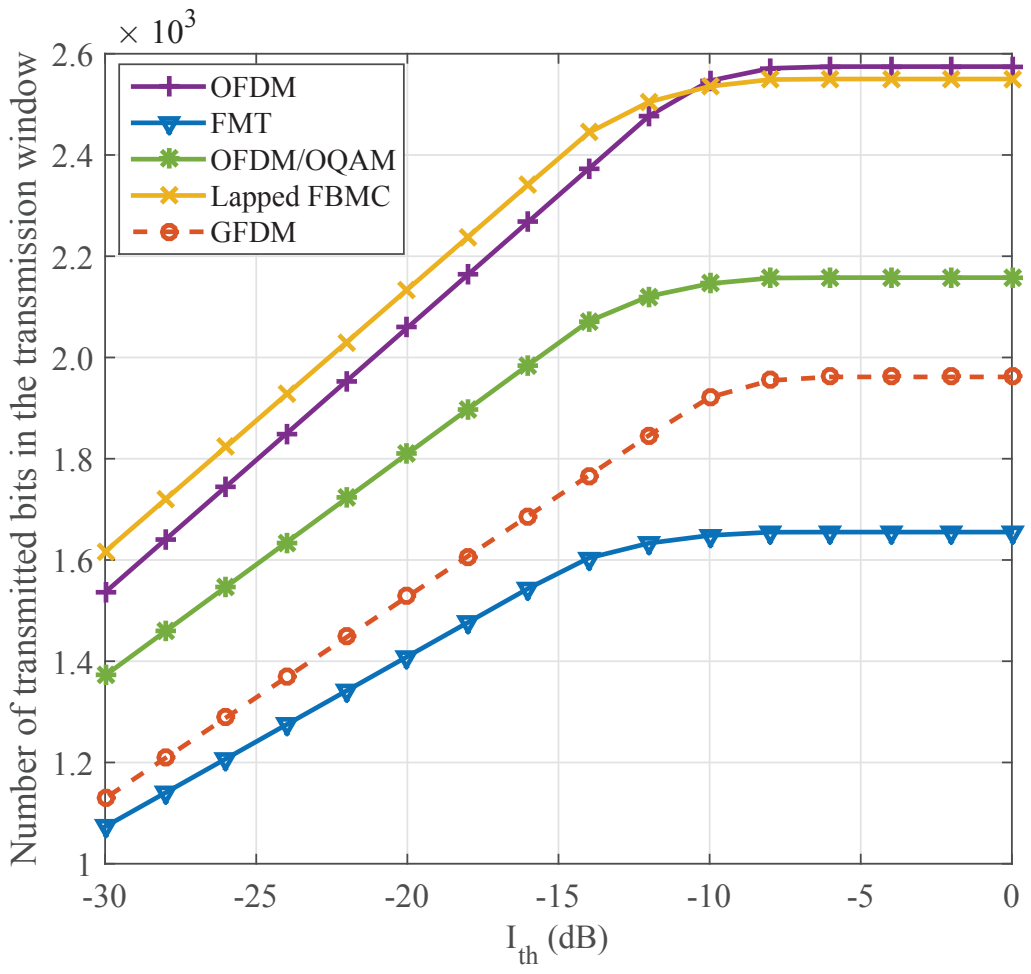

Fig. 7: Bits transmitted by different waveforms in 12 subcarriers during one TTI in function of the interference threshold.

## VI. Conclusion

In this study, we investigated a scenario where a D2D pair coexists with an OFDM based incumbent network. The D2D pair adopts an alternative 5G waveform to OFDM. Time-frequency misalignment of the D2D was taken into account to generate the interference tables from different waveforms to an OFDM receiver. This is in contrast to the usual analysis available in the literature, where the same waveform is considered for the source and the victim of interference. Through numerical results, we have shown that it is not worth synchronizing the D2D pair in time domain with respect to the incumbent network. Interference tables derived in this paper allowed us to analyse the maximum rate achievable by the D2D pair under different interference constraints. We have shown that the communication window size has a direct impact on the efficacy of the waveform utilized by the D2D pair. For short D2D transmission windows, OFDM, GFDM and Lapped FBMC seem to be appropriate candidates. We also showed that under stringent interference constraints and wide transmission windows, OFDM/OQAM and Lapped FBMC are strong candidates.